# Advancements in High Density and Fast Scintillation Detector Materials

R. Hawrami[1], E. Ariesanti[2], A. Burger[2], and H. Parkhe[2]

[1]Xtallized Intelligence, Inc., Nashville, TN 37211

[2]Fisk University, Nashville, TN 37208

**Abstract**

Nuclear and high energy physics research has a need for new, high performance scintillators with high light yields; high densities, fast decay times, and are radiation hard. In this paper we present crystal growth and results from 16-mm diameter cerium (Ce)-doped $Tl_2LaCl_5$ (TLC) and europium (Eu)-doped $TlCa_2Br_5$ (TCB) as well as one-inch diameter cerium-doped $Tl_2GdBr_5$ (TGB) and europium-doped $TlSr_2I_5$ (TSI), each grown in a two-zone vertical furnace by the modified Bridgman method. Samples extracted and processed from the grown boule are characterized for their scintillation properties like energy resolution, light yield, decay time and non-proportionality. Energy resolution (FWHM) at 662 keV of 5.1%, 3.4%, 4.0%, and 3.3% are obtained for samples of TGB, TLC, TCB, and TSI, respectively. Ce-doped TGB and TLC have single decay time components of 26 ns and 48 ns, respectively, while Eu-doped TCB and TSI have long decay times with primary decay constants of 571 ns and 630 ns. These compounds exhibit good proportionality behavior when compared to NaI:Tl and BGO.

**Index Terms**

Crystal growth, Gamma-ray detector, Scintillation detector, Thallium-based metal halide crystals.

## I. INTRODUCTION

The field of inorganic scintillators has expanded in the last three decades, with the (re)discovery and successful growth of novel and advanced scintillation compounds. Demand for high light yield, high density, and fast scintillators necessitate a continuous search for new materials. Traditional scintillators such as Tl-doped sodium iodide (NaI) and cesium iodide (CsI) have been very reliable standards, supported by decades of research and proven performance. However, various new applications require bright materials that also have high densities and fast decay times. For nearly two decades emerging new scintillators such as rare-earth binary compounds of $CeX_3$[1, 2] and $LaX_3$ [3],as well as ternary metal halide compounds of $Cs_2AX_5$[4], where A = La or Ce, and X = Cl, Br, or I (halides), have demonstrated the potentials of these metal halides as next-generation scintillation detectors. Rediscovered Eu-doped $SrI_2$, with a light yield as high as 110,000 ph/MeV and moderate density of 4.55 g/cm$^3$, has also shown the potential of alkaline metal halide scintillators [5, 6].Currently used inorganic scintillators for high energy physics experiments include LSO, $PbWO_4$, CsI, and $LaBr_3$. These scintillators have some, if not all, of the desired properties such as high densities, high light yields, and fast decay times.The combination of high–light yield and fast response can be found in $Ce^{3+}$, $Pr^{3+}$, or $Nd^{3+}$ -doped lanthanide scintillators, while the maximum light yield conversion of 100,000 photons/MeV can be found in $Eu^{3+}$-doped $SrI_2$. However, growth of those oxide-based and lanthanide-doped scintillators is inefficient and expensive because it requires high-temperature furnaces [4-7]. Along with most binary halides including $LaBr_3$:Ce, NaI:Tl and elpasolites crystals are characterized by their relatively low detection efficiency. This is primarily because these materials do not contain any constituents with very high Z and their density is moderate. This not only impacts their overall photon stopping power but also compromises the photo-fraction (defined as the ratio of photoelectric cross-section to total cross-section) for detection of low-energy photons. Higher photo-fraction is important, because it provides more counts in the desired photopeak region of the energy spectrum, making the task of isotope identification easier and faster [8].

Recently high detection efficiency Tl-based scintillation crystals have attracted good attention from worldwide scintillator researchers. These compounds have been investigated and very promising initial results have been published, for example Ce-doped $Tl_2LaCl_5$ (TLC) [7] as well as *intrinsic* (i.e., undoped) $TlMgCl_3$ (TMC) and $TlCaI_3$ (TCI) [8,9]. These new compounds are of high densities (> 5 g/cm$^3$), bright (light yields between 31,000 and 76,000 ph/MeV for 662 keV photons), fast decay times (36 ns (89%) for TLC; 46 ns (9%) for TMC; 62 ns (13%) for TCI), and moderate melting points (between 500 and 700°C). As seen further in the published results, Tl-based scintillators such as the ones previously mentioned have promising properties desirable for high energy physics as well as homeland security applications [7-11]. In this paper we are reporting on the growth and scintillation characterization ($^{137}$Cs spectra, decay times and non-proportionality behavior) of TGB, TLC, TCB and TSI crystals.

## II. EXPERIMENTAL PROCEDURE

Based on stoichiometric calculations, the appropriate amounts of starting halide compounds (all in powder form with 4N purity) were loaded into a growth ampoule: for $Tl_2GdBr_5$ (TGB) is mix of 2TlBr+ $GdBr_3$ with 3%$CeCl_3$ as dopant; for $Tl_2LaCl_5$ (TLC) is the mix of 2TlCl+ $LaCl_3$ with 3%$CeCl_3$, as dopant; for $TlCd_2Br_5$(TCB) is the mix TlBr + 2$CaBr_2$ with 5% $EuBr_2$ as dopant; for $TlSr_2I_5$(TSI) is mix of TlI + 2$SrI_2$ with 5% $EuI_2$ as dopant. TGB and TSI growth runs were conducted in ∅1-inch (inner diameter) quartz ampoules, while TLC and TCB growth runs in ∅16-mm (inner diameter) quartz ampoules. After material loading each ampoule was subsequently sealed under high vacuum 2.4x10$^{-5}$Torr and placed in a two-zone vertical Bridgman furnace. Furnace zone temperatures were set such that the temperature profiles would facilitate melting at around 780°C, 680°C, 550°C, and 630°C for TGB, TLC, TCB and TSI, respectively. For each experiment, the crystal growth commenced at a rate of 15-20 mm/day and the post-crystallization cooling at a rate of 100°C to 150°C/day.

After the cooling down procedure was completed, the ampoules were retrieved from the furnaces and samples were harvested from the boule. Each sample was lapped and polished with sandpapers. Mineral oil was used for lubrication during processing as well as for sample protection from moisture because TLC, TGB, TCB and TSI were hygroscopic. The polished samples were tested for their radiometric and scintillation properties. To measure energy resolution and non-proportionality behavior, each sample was placed in mineral oil in a quartz cup wrapped with Teflon tape as a reflector. A piece of Gore ® flexible Teflon sheet was be used as the back reflector. Using BC-630 optical grease, the oil cup was coupled to aR6231-100 Hamamatsu ∅2-inchsuper bi-alkali photomultiplier tube (PMT). The signals from the anode were fed to a Canberra 2005 preamplifier, a Canberra 2020 amplifier, and a MCA8000D multi-channel analyzer. $^{137}$Cs spectra were collected by each sample and energy resolution (FWHM) at the full energy peak of 662 keV. Spectra from other gamma-ray check sources, including $^{22}$Na, $^{57}$Co, $^{60}$Co, $^{133}$Ba, $^{152}$Eu, and $^{241}$Am, were also collected to determine the non-proportionality data for each compound. To obtain luminescence decay time information, each sample was irradiated with a $^{137}$Cs check source and the signal waveforms collected at the PMT anode were recorded with CAEN DT5720C digitizer and then were analyzed offline.

## III. Results and Analysis

### A. Crystal Growth and Sample Processing

Samples from successfully grown crystal boules of ∅1-inch TGB, ∅16-mm TLC, ∅16-mm TCB, and ∅1-inch TSI are shown in Fig. 1: 10×25×30 mm³ TGB, ∅16-mm × 16 mm TLC, ∅16-mm × 8 mm TCB, and 12×15×20 mm³ TSI. Each sample was processed immediately prior to characterization with previously described processing procedures.

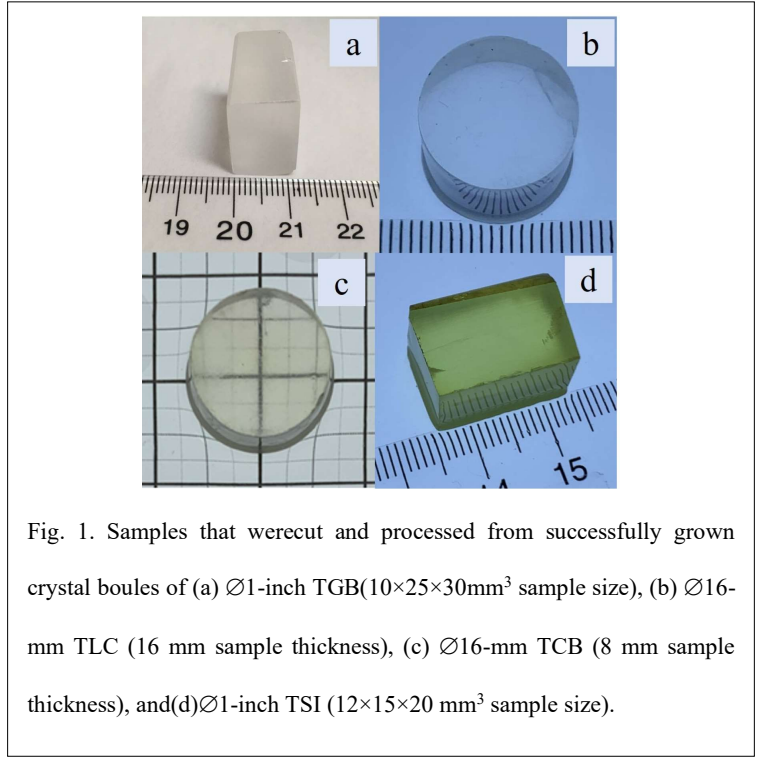

Fig. 1. Samples that werecut and processed from successfully grown crystal boules of (a) ∅1-inch TGB(10×25×30mm³ sample size), (b) ∅16-mm TLC (16 mm sample thickness), (c) ∅16-mm TCB (8 mm sample thickness), and(d)∅1-inch TSI (12×15×20 mm³ sample size).

### B. Scintillation Detector Performance

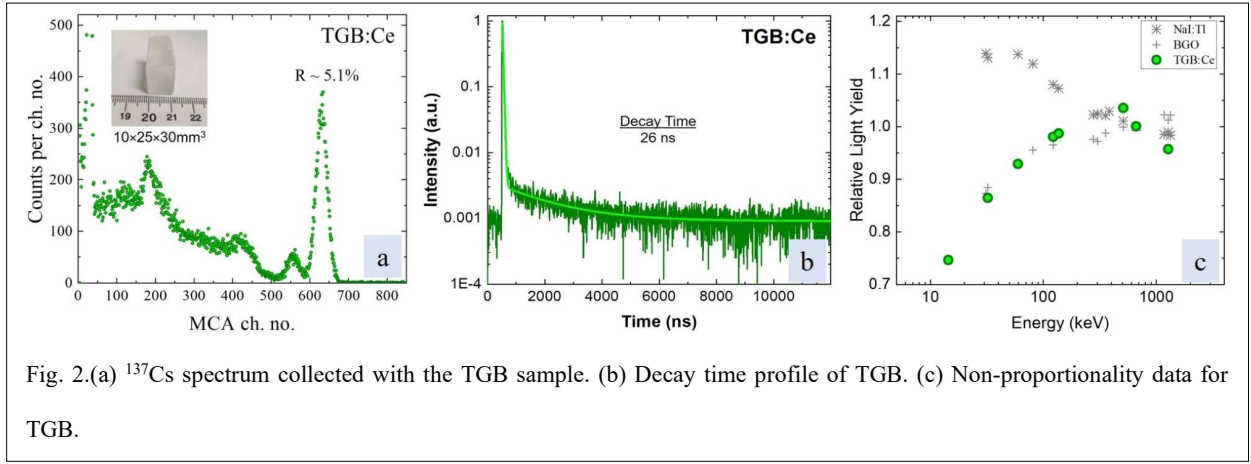

Fig. 2.(a) $^{137}$Cs spectrum collected with the TGB sample. (b) Decay time profile of TGB. (c) Non-proportionality data for TGB.

Results for Ce-doped TGB and TLC are shown in Figs. 2 and 3, respectively. Energy resolution of 5.1% (FWHM) at 662 keV was calculated for TGB (Fig. 2(a)) while energy resolution of 3.4% (FWHM) was calculated for TLC (Fig. 3(a)). Ce-doped scintillators are known to produce fast decay times [7], as seen in Figs. 2(b) and 3(b), where single decay time constant of 26 ns for TGB and 48 ns was measured for TLC. The proportionality behavior or the relative light yield data as a function of photon energy for each scintillator is shown in Figs. 2(c) and 3(c), respectively. For either sample less than linear response (outside of ±5% from unity) was observed for γ-ray energy less than 200 keV. The reasons are yet to be determined. Although the samples were characterized in oil to avoid moisture interaction, nevertheless, slight degradation might have occurred, thus decreasing the apparent light yield.

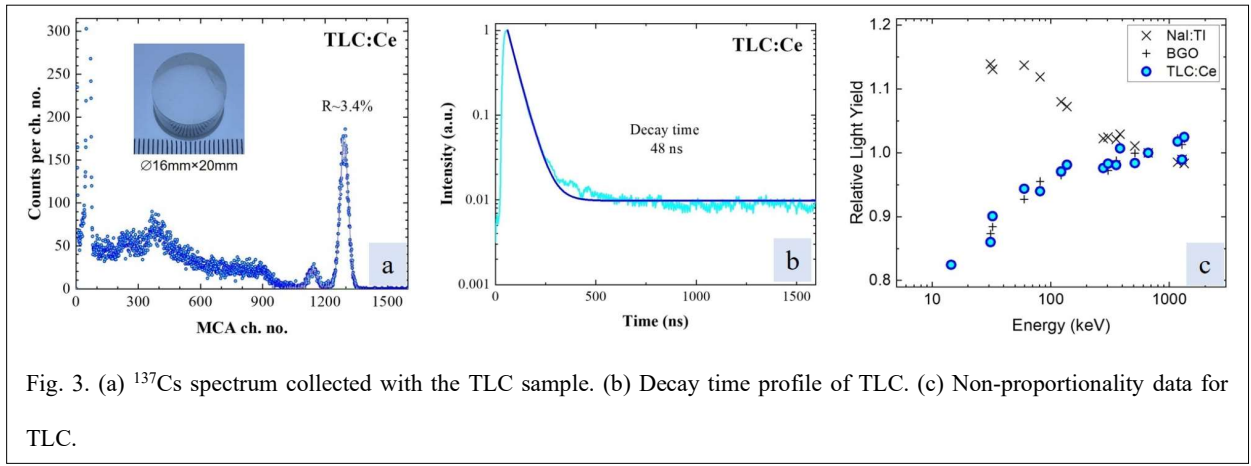

Fig. 3. (a) $^{137}$Cs spectrum collected with the TLC sample. (b) Decay time profile of TLC. (c) Non-proportionality data for TLC.

Results for Eu-doped TGB and TSI are shown in Figs. 4 and 5, respectively. Energy resolution of 4.0% (FWHM) at 662 keV was calculated for TGB (Fig. 4(a)) while energy resolution of 3.3% (FWHM) was calculated for TSI (Fig. 5(a)). Eu-doped scintillators are known to have long decay times [11, 12], as seen in Figs. 4(b) and 5(b). The decay time profile for TCB (Fig. 4(b)) was fitted with three exponential functions, resulting in decay constants of 541 ns (70%), 973 ns (9%), and 3.3 μs (21%). The decay time profile for TSI (Fig. 5(a)) was fitted with two exponential functions, resulting in decay constants of 630 ns (74%) and 3.6 μs (26%). The proportionality behavior or the relative light yield data as a function of photon energy for each scintillator is shown in Figs. 4(c) and 5(c), respectively. For either sample less than linear response (outside of ±5% from unity) was observed for γ-ray energy less than 50 keV. The reasons are yet to be determined. Similar with the sample treatment for TGB and TLC, although the TGB and TSI samples were characterized in oil to avoid moisture interaction, nevertheless, slight degradation might have occurred, thus decreasing the apparent light yield.

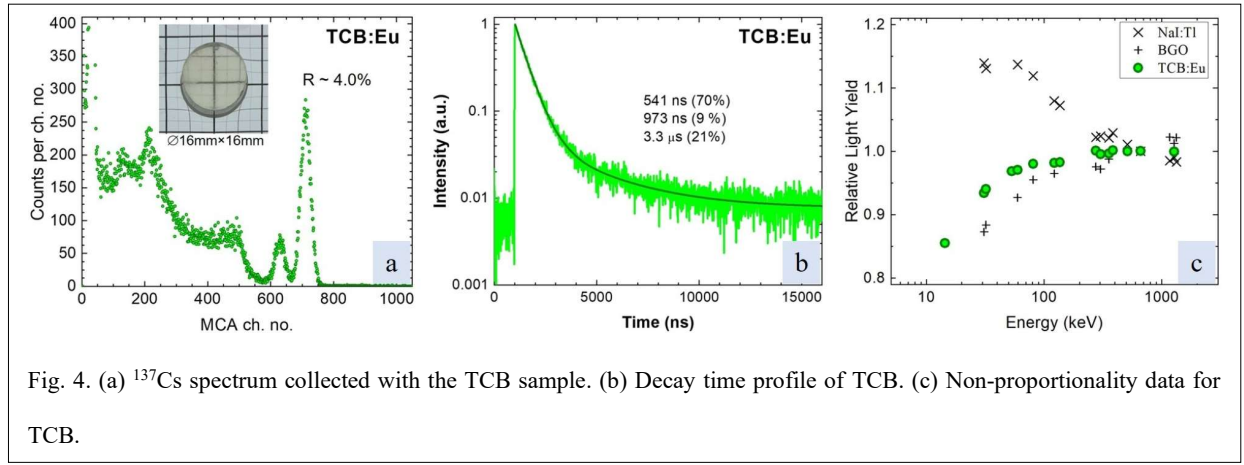

Fig. 4. (a) $^{137}$Cs spectrum collected with the TCB sample. (b) Decay time profile of TCB. (c) Non-proportionality data for TCB.

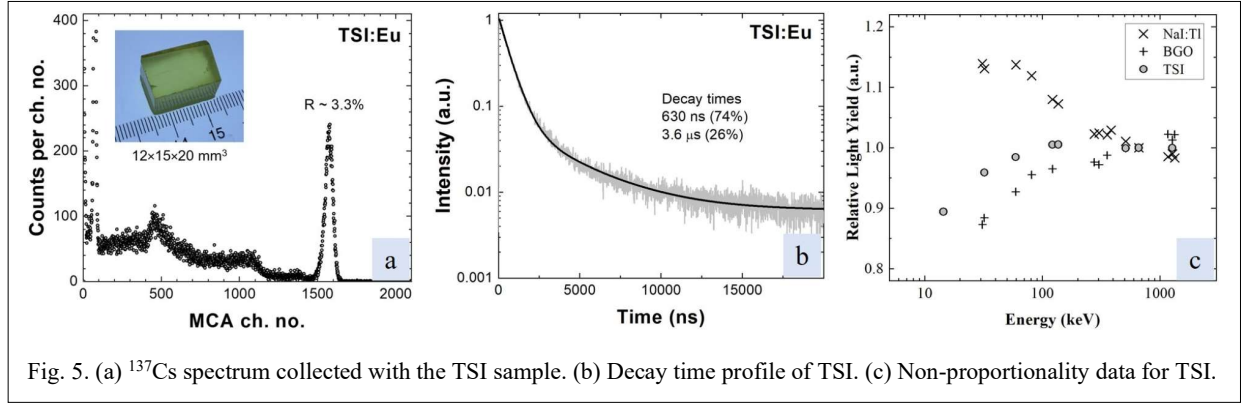

Fig. 5. (a) $^{137}$Cs spectrum collected with the TSI sample. (b) Decay time profile of TSI. (c) Non-proportionality data for TSI.

## IV. Conclusions

This paper reports on successful growth runs and scintillating performance of 16-mm diameter cerium-doped $Tl_2LaCl_5$ (TLC) and europium-doped $TlCa_2Br_5$ (TCB) as well as one-inch diameter cerium-doped $Tl_2GdBr_5$ (TGB) and $TlSr_2I_5$ (TSI), all grown by the vertical Bridgman method. Samples were extracted from the boules and processed for characterization. Energy resolution of 5.1%, 3.4%, 4.0%, and 3.3% are obtained for samples of TGB, TLC, TCB, and TSI, respectively. Ce-doped TGB and TLC have single decay time components of 26 ns and 48 ns, respectively, while Eu-doped TCB and TSI have long decay times with primary decay constants of 571 ns and 630 ns. These compounds exhibit good proportionality behavior when compared to NaI:Tl and BGO.

## V. Acknowledgments

The work in this paper was partly supported by personal funds from Xtallized Intelligence, Inc. personnel.